\documentclass[sigconf]{acmart}
\AtBeginDocument{%
  \providecommand\BibTeX{{%
    \normalfont B\kern-0.5em{\scshape i\kern-0.25em b}\kern-0.8em\TeX}}}

\copyrightyear{2023}
\acmYear{2023}
\setcopyright{acmcopyright}
\acmConference[KDD '23]{Proceedings of the 29th ACM SIGKDD Conference on Knowledge Discovery and Data Mining}{August 6--10, 2023}{Long Beach, CA, USA}
\acmBooktitle{Proceedings of the 29th ACM SIGKDD Conference on Knowledge Discovery and Data Mining (KDD '23), August 6--10, 2023, Long Beach, CA, USA}
\acmPrice{15.00}
\acmISBN{979-8-4007-0103-0/23/08}
\acmDOI{10.1145/3580305.3599779}

\usepackage{amsmath,amsfonts,bm}

\usepackage{IEEEtrantools}

\usepackage{amsthm}
\theoremstyle{definition} \newtheorem{definition}{Definition}

\usepackage{siunitx}

\usepackage{graphicx}

\usepackage{caption}
\usepackage{subcaption}

\begin{document}

\title{Ball Trajectory Inference from Multi-Agent Sports Contexts Using Set Transformer and Hierarchical Bi-LSTM}

\author{Hyunsung Kim}
\orcid{0000-0002-6286-5160}
\affiliation{%
  \institution{Fitogether Inc.}
  \city{Seoul}
  \country{South Korea}
}
\email{hyunsung.kim@fitogether.com}

\author{Han-Jun Choi}
\affiliation{%
  \institution{Kangwon National University}
  \city{Chuncheon}
  \country{South Korea}
}
\email{hanjunchoi@kangwon.ac.kr}

\author{Chang Jo Kim}
\affiliation{%
  \institution{Fitogether Inc.}
  \city{Seoul}
  \country{South Korea}
}
\email{changjo.kim@fitogether.com}

\author{Jinsung Yoon}
\affiliation{%
  \institution{Fitogether Inc.}
  \city{Seoul}
  \country{South Korea}
}
\email{jinsung.yoon@fitogether.com}

\author{Sang-Ki Ko}
\orcid{0000-0002-5406-5104}
\affiliation{%
  \institution{Kangwon National University}
  \city{Chuncheon}
  \country{South Korea}
}
\additionalaffiliation{%
  \institution{Fitogether Inc.}
  \city{Seoul}
  \country{South Korea}
}
\email{sangkiko@kangwon.ac.kr}

\renewcommand{\shortauthors}{Hyunsung Kim et al.}

\begin{abstract}
As artificial intelligence spreads out to numerous fields, the application of AI to sports analytics is also in the spotlight. However, one of the major challenges is the difficulty of automated acquisition of continuous movement data during sports matches. In particular, it is a conundrum to reliably track a tiny ball on a wide soccer pitch with obstacles such as occlusion and imitations. Tackling the problem, this paper proposes an inference framework of ball trajectory from player trajectories as a cost-efficient alternative to ball tracking. We combine Set Transformers to get permutation-invariant and equivariant representations of the multi-agent contexts with a hierarchical architecture that intermediately predicts the player ball possession to support the final trajectory inference. Also, we introduce the reality loss term and postprocessing to secure the estimated trajectories to be physically realistic. The experimental results show that our model provides natural and accurate trajectories as well as admissible player ball possession at the same time. Lastly, we suggest several practical applications of our framework including missing trajectory imputation, semi-automated pass annotation, automated zoom-in for match broadcasting, and calculating possession-wise running performance metrics.
\end{abstract}





\maketitle

\section{Introduction}

With the rapid progress in artificial intelligence (AI) and machine learning, there also is a growing interest in automated sports analytics for providing a competitive advantage to teams or individual players~\cite{Decroos2020}. However, one of the major challenges is still the data collection process as the huge amount of high-quality data is necessary to apply advanced machine learning techniques for sophisticated analyses. In competitive team sports such as soccer, basketball, ice hockey, and so on, there are an enormous amount of accessible video data including broadcast videos, but it is very difficult to extract the essential information such as the trajectories of players and the ball from these videos.

Especially, ball tracking is a critical problem for video-based analysis~\cite{Kamble2018, Kamble2019a, Kamble2019b} in team sports. However, it is well known to be very difficult to reliably track the ball from videos due to the small size of the ball and the occlusion problem. Wang et al.~\cite{Wang2014} proposed a novel approach by formulating a tracking algorithm in terms of deciding who has the ball at a given time. Maksai et al.~\cite{Maksai2016} introduced a more principled approach by modeling the interaction between the ball and the players and even the physical constraints of ball trajectories for better ball tracking performance on soccer videos. On the industry side, FIFA recently introduced the semi-automated offside technology~\cite{FIFA2022} at the World Cup Qatar 2022 with automatic ball tracking by placing an inertial measurement unit (IMU) sensor inside the ball, but it is not broadly applicable due to a very expensive cost~\cite{Price2022}.

Compared to ball tracking, it is relatively easier to acquire player trajectories as players are much bigger than a ball in videos. Moreover, there are different types of systems for tracking players in team sports other than video-based systems such as global positioning systems (GPS) or local positioning systems (LPS). Currently, there are various data providers of player tracking data, either using video-based tracking~\cite{StatsPerform,SecondSpectrum,ChyronHego,MetricaSports}, GPS-based tracking~\cite{Catapult,STATSports,Fitogether}, or LPS-based tracking~\cite{Kinexon}. Some video-based data providers are acquiring ball tracking data by first collecting match events and combining it with player tracking data~\cite{OptaVision}, but it still needs a lot of manual work for event annotation.

In this paper, we propose a framework for inferring the ball trajectories from player trajectories instead of directly tracking them from sports videos. We implement permutation-invariant and equivariant encoders using Set Transformers~\cite{Lee2019} to represent multi-agent contexts of sports games. Inspired by Zhan et al.~\cite{Zhan2019}, we combine these context encoders with a hierarchical recurrence structure that intermediately predicts the domain-specific semantics such as player-level ball possession to enhance the final prediction performance. Additionally, we introduce the reality loss term and postprocessing to secure the estimated trajectories to be physically realistic in that they only change direction when they are carried or kicked by players. The experimental results show that our method predicts accurate ball trajectories with a mean position error smaller than \SI{3.7}{\meter} while predicting player ball possession with an accuracy of \SI{64.7}{\%} as an intermediate product.

The main contribution of our study is that we give shape to a new way of ball data collection in sports, neither relying on heavy camera infrastructure nor hard manual work but based on machine learning techniques and player tracking data that are relatively easy to obtain. Moreover, it enables semi-automated event annotation by detecting ball-related events so that humans only need to correct errors. We expect that the proposed method would contribute to the sports industry by lowering the cost of data acquisition and eventually the entry barrier for the sports analytics ecosystem.

\section{Related Work}

\subsection{Ball Trajectory Inference from Player Trajectories in Team Sports}
Though ball tracking from sports videos is a topic of interest in computer vision, only a few studies~\cite{Amirli2022,Gongora2021} tried to estimate ball trajectories not relying on optical tracking but only using players' movement data. Amirli et al.~\cite{Amirli2022} aggregated players' locations and speeds to make handcrafted input features and constructed a neural network regressor to estimate ball locations. However, they did not employ sophisticated architectures to encode the sequential nature or permutation-invariance of multi-agent trajectories. As a result, their framework shows too large position errors (\SI{7.56}{m} along the x-axis and \SI{5.01}{m} along the y-axis) to be utilized in practice. On the other hand, Gongora~\cite{Gongora2021} adopted Set Transformer~\cite{Lee2019} to encode permutation-invariant game contexts and used sequence models such as Transformer~\cite{Vaswani2017} or InceptionTime~\cite{Fawaz2020} to predict ball trajectories. The difference of ours from this approach is that we build a hierarchical architecture and employ another type of sequence model (i.e., Bi-LSTM~\cite{Hochreiter1997}). In Section~\ref{se:experiments}, we demonstrate that the use of a hierarchical architecture and Bi-LSTMs both promote more accurate ball trajectory prediction and explain why RNN models outperform Transformer in this problem.

\subsection{Multi-Agent Trajectory Prediction}

Predicting the future trajectories of objects is a crucial task, especially for autonomous platforms like self-driving cars or social robots. Zhang et al.~\cite{Zhang2020} proposed the end-to-end Spatio-Temporal-Interactive Network (STINet) to model pedestrians. Felsen et al.~\cite{Felsen2018} constructed a Conditional Variational Autoencoder (CVAE) for predicting adversarial multi-agent motion. Yeh et al.~\cite{Yeh2019} proposed the Graph Variational RNN (GVRNN) using graph structure for the permutation-equivariant representation of multi-agent trajectories in sports games. Zhan et al.~\cite{Zhan2019} introduced a hierarchical architecture that first predicts the long-term intents of agents and then generates future trajectories conditioned on the intents.

Meanwhile, the problem of missing trajectory imputation is also a significant topic in spatiotemporal data mining. BRITS~\cite{Cao2018} (Bidirectional Recurrent Imputation for Time Series) is a method based on bidirectional RNNs for missing value imputation in time series data. However, BRITS cannot resolve the error propagation problem when imputing long-range sequences due to its autoregressive nature. In order to better handle the problem, NAOMI~\cite{Liu2019} (Non-Autoregressive Multiresolution Imputation) exploits the multiresolution structure that decodes recursively from coarse to fine-grained resolutions using a divide-and-conquer strategy. Also, Qi et al.~\cite{Qi2020} proposed an imitative non-autoregressive modeling method to simultaneously handle the trajectory prediction task and the missing value imputation task. Omidshafiei et al.~\cite{Omidshafiei2022} introduced Graph Imputer for predicting players' off-screen behavior in soccer, where the model is similar to GVRNN in that it combines graph networks and the Variational RNN~\cite{Chung2015}, except for using a bidirectional structure since it can observe partial future trajectories.

Our ball trajectory inference task is closely related to the aforementioned problems, but there is a clear difference in the detailed setup. The goal of the original trajectory prediction or imputation problem is to predict the behavior of intelligent agents in unknown frames, given the information including the target agent's partial trajectories. In contrast, we have to estimate the entire trajectory of the ball that does not have an intent, given the trajectories of other agents that affect the target trajectories. Therefore, believing that our problem is deterministic rather than stochastic, we construct an LSTM-based regression model instead of generative models that many other trajectory prediction frameworks are based on.

\begin{figure*}
  \includegraphics[width=0.98\textwidth]{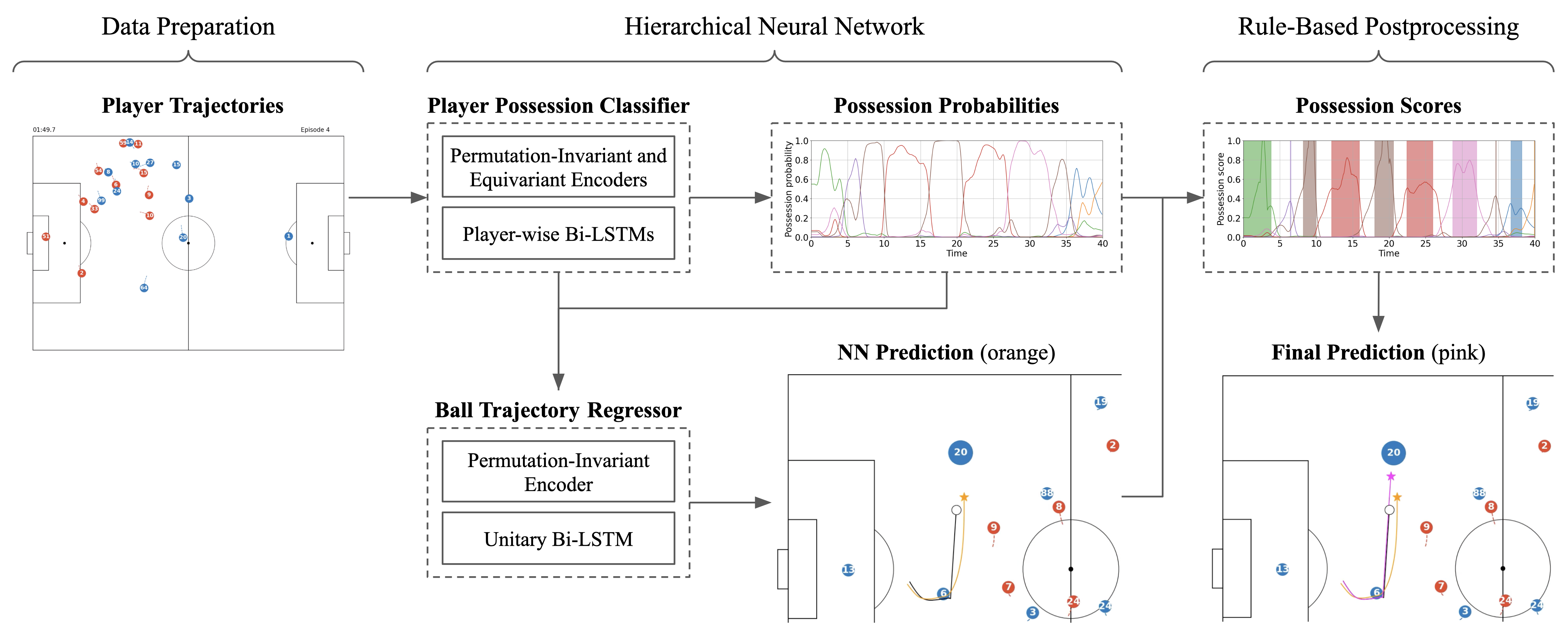}
  \caption{Overview of the proposed ball trajectory inference framework.}
  \label{fig:overview}
\end{figure*}

\section{Learning Approach}

In Section~\ref{se:problem_def}, we formally define the considered problem. Section~\ref{se:player_poss} and Section~\ref{se:pi_pe_encoders} explain the two notable points of the neural network part of our framework, where we elaborate on the details in Section~\ref{se:framework}. Section~\ref{se:loss_function} describes the loss function for training the network, and Section~\ref{se:postprocess} introduces the rule-based postprocessing algorithm to enforce the output trajectories to be realistic.

\subsection{Problem Formulation} \label{se:problem_def}

Our ultimate goal is to find the ball conditioned on the player locations. However, since we construct a hierarchical framework that also predicts player-level ball possession, our approach actually solves the following two problems at once.
\begin{itemize}
    \item \textbf{Ball possessor prediction:} Given a set $X_{1:T}^{P} = \{ \mathbf{x}_{1:T}^p \}_{p \in P}$ of player trajectories, find the player $q_t$ that possesses the ball at each time $t$ where $1 \le t \le T$. (We describe the definition of player possession in Section~\ref{se:player_poss}.)
    \item \textbf{Ball trajectory prediction:} Given a set $X_{1:T}^{P} = \{ \mathbf{x}_{1:T}^p \}_{p \in P}$ of player trajectories, find the ball trajectory $\mathbf{y}_{1:T}$.
\end{itemize}

\subsection{Player-Level Ball Possession as an Intermediate Target Variable}
\label{se:player_poss}

Zhan et al.~\cite{Zhan2019} proposed a hierarchical VRNN framework for future trajectory generation of multiple agents, using intermediate weak labels that capture macroscopic behavioral semantics. Inspired by the study, we also design a hierarchical Bi-LSTM~\cite{Hochreiter1997} structure with intermediate labels to enhance the final prediction performance.

In our study, we utilize player-level ball possession as an intermediate target variable. Namely, the model first produces $\hat{\mathbf{g}}_t = (\hat{g}_t^{p_1}, \ldots , \hat{g}_t^{p_N})$ where $\hat{g}_t^{p_i}$ is the probability that the player $p_i$ possesses the ball at $t$, and predicts $\hat{\mathbf{y}}_t$ conditioned on $\hat{\mathbf{g}}_t$. Since the ball is always either controlled by a player, in transition from one player to another, or out of play, this information about player ball possession can be a strong hint for final trajectory inference.

Here we define that a player ``possesses'' the ball if the player is controlling the ball or he or she is the next controller of the ball. That is, we label that the ball possession changes from player A to B right after A passes the ball to B. The reason why we assign a specific player even when the ball is in transition is that it causes a significant class imbalance if we label the possession of those 
timesteps as ``void'' across the board.

\subsection{Permutation-Invariant and Equivariant Representations of Sports Contexts}
\label{se:pi_pe_encoders}

An overall game context of competitive team sports is \emph{partially permutation-invariant}. That is, the order of players in each team is permutable when representing a game context, while players in a team are not exchangeable with those in the other team. On the other hand, player-level tasks such as player trajectory prediction or player ball possession inference need to be \emph{permutation-equivariant} in that a permutation of the input players leads to the same permutation at the output.

To this end, we adopt the Set Transformer~\cite{Lee2019}, a state-of-the-art methodology for permutation-invariant or equivariant embedding. It consists of an encoder block and a decoder block where the encoder produces a permutation-equivariant embedding of the set input, and the decoder returns a permutation-invariant embedding of a fixed dimension. Hence, we use a whole structure of a Set Transformer for permutation-invariant representation of game contexts, while only taking the encoder part of a Set Transformer for permutation-equivariant representation.

In our study, the ball trajectory prediction is a partially permutation-invariant (PPI) task and the ball possessor prediction is a partially permutation-equivariant (PPE) one. For the former, we construct a PPI encoder consisting of Set Transformers per team. For the latter, on the other hand, we additionally construct fully permutation-equivariant (FPE) and full permutation-invariant (FPI) encoders as well as a PPE encoder and combine them to improve the performance. We explain the detailed architectures of these encoders in Section~\ref{se:framework} and the reasons for adopting each part in Appendix~\ref{se:ablation_encoders}.

\subsection{Detailed Hierarchical Framework} \label{se:framework}

\begin{figure}[t!]
    \centering
    \begin{subfigure}[t]{0.45\textwidth}
        \includegraphics[width=\textwidth]{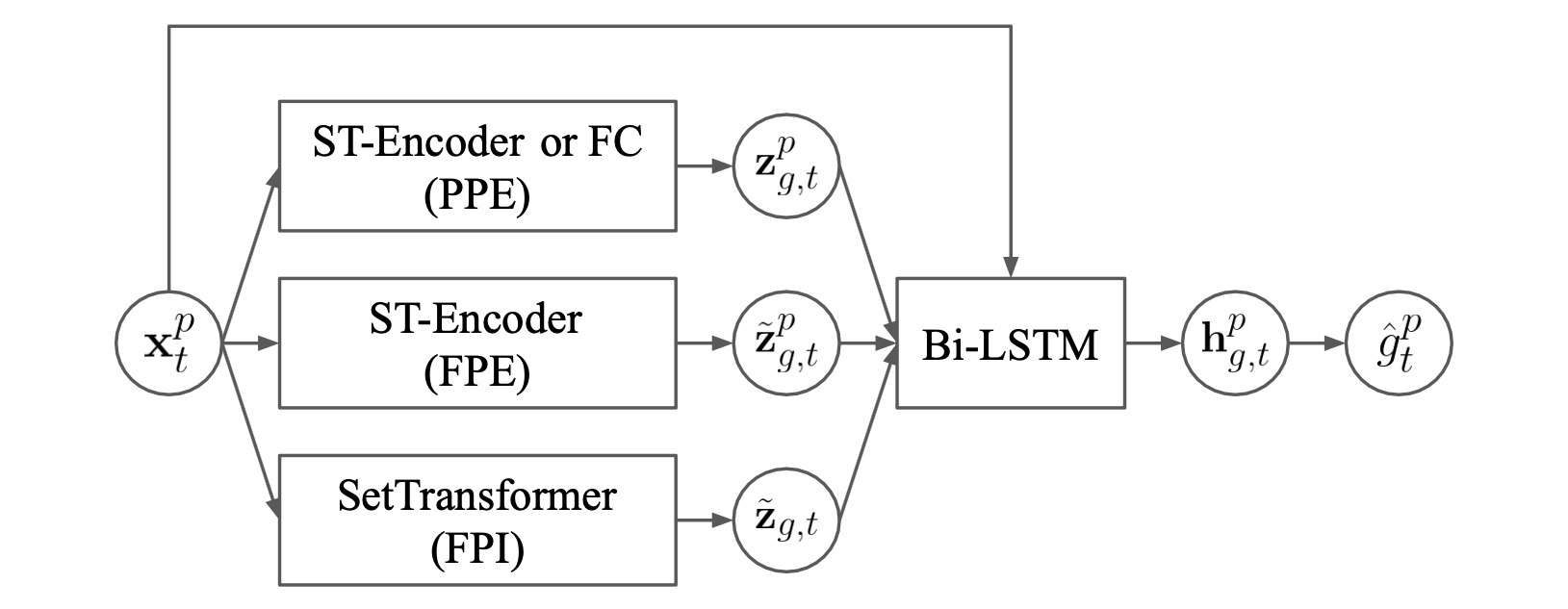}
        \caption{Player Possession Classifier.}
        \label{fig:model_ppc}
    \end{subfigure}
    \begin{subfigure}[t]{0.45\textwidth}
        \centering
        \includegraphics[width=\textwidth]{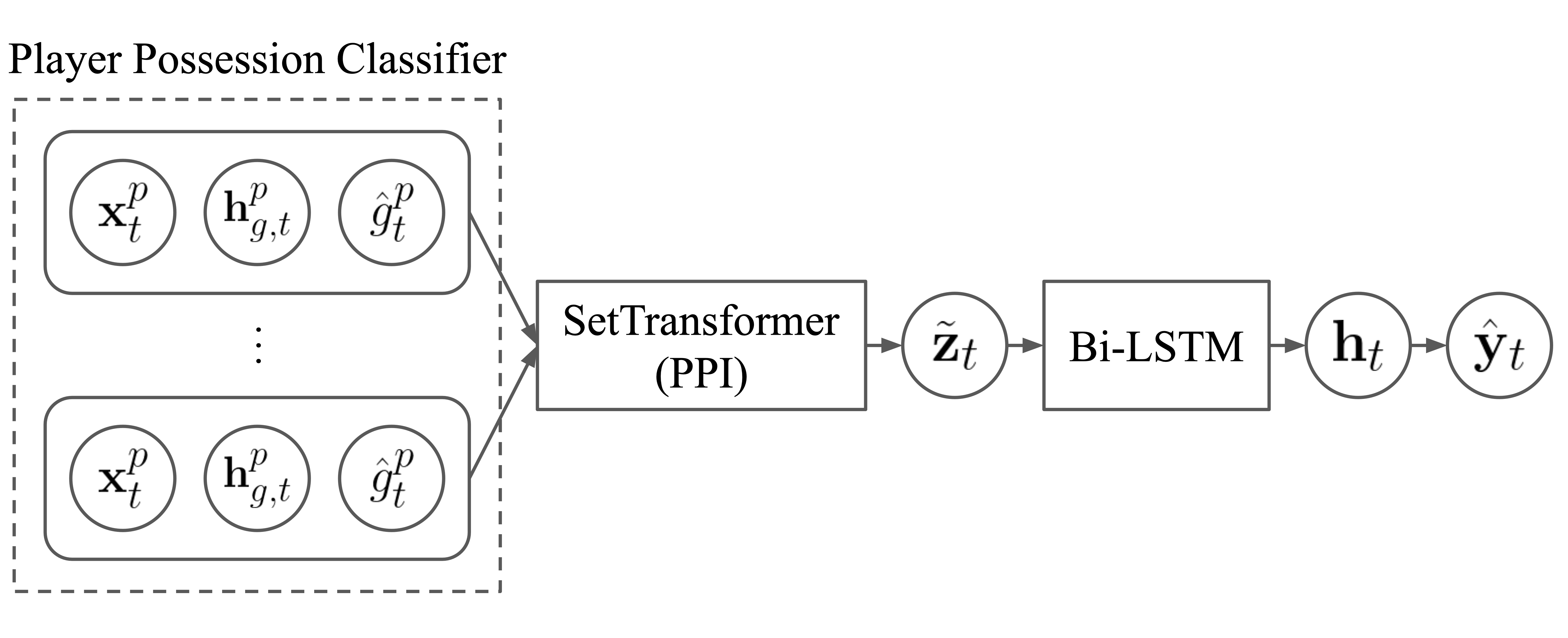}
        \caption{Ball Trajectory Regressor.}
        \label{fig:model_btr}
    \end{subfigure}
    \caption{Architectures of PPC and BTR.}
    \label{fig:architecture}
\end{figure}

In this section, we elaborate on our hierarchical inference framework. The model consists of \emph{Player Possession Classifier} (PPC) that estimates players' ball possession probabilities $\mathbf{g}_{1:T}$ and \emph{Ball Trajectory Regressor} (BTR) that finds the ball trajectory $\mathbf{y}_{1:T}$ using the information from PPC. Figure~\ref{fig:architecture} depicts the model architectures.

\subsubsection{Player Possession Classifier} \label{se:model_ppc}
First, we convert input features into context-aware player representations in different ways by partially permutation-equivariant encoding (PPE), fully permutation-equivariant encoding (FPE), and fully permutation-invariant encoding (FPI). In Appendix~\ref{se:ablation_encoders}, we elucidate why these three variants of game context encoder are needed by presenting intuitive reasons and carrying out an ablation study.

For PPE encoding, we employ a Set Transformer encoder (denoted as ST-Encoder below) for input player features $\{ \mathbf{x}_t^p \}_{p \in P_k}$ in each team $P_k$ ($k = 1, 2$) at time $t$ to get teammate-aware player embeddings $\{ \mathbf{z}_t^p \}_{p \in P_k}$ as follows:
\begin{IEEEeqnarray}{rCl}
    (\mathbf{z}_{g,t}^{p_1}, \ldots, \mathbf{z}_{g,t}^{p_n})        & = & \text{ST-Encoder} (\mathbf{x}_t^{p_1}, \ldots, \mathbf{x}_t^{p_n}) \label{eq:ppe_team1} \\
    (\mathbf{z}_{g,t}^{p_{n+1}}, \ldots, \mathbf{z}_{g,t}^{p_{2n}}) & = & \text{ST-Encoder} (\mathbf{x}_t^{p_{n+1}}, \ldots, \mathbf{x}_t^{p_{2n}}) \label{eq:ppe_team2} \\
    \mathbf{z}_{g,t}^{p_{2n+i}}  & = & \text{FC} (\mathbf{x}_t^{p_{2n+i}}),\quad i = 1, \ldots, 4 \label{eq:ppe_outsides}
\end{IEEEeqnarray}
where $\{ p_1, \ldots, p_n \}$ and $\{ p_{n+1}, \ldots, p_{2n} \}$ are the players in team $P_1$ and $P_2$, respectively, and $P_0 = \{ p_{2n+i} \}_{i=1}^4$ are states of the ball out of the four pitch lines. We set the input coordinates $\mathbf{x}_t^{p_{2n+i}}$ for these ball-out states as the middle points of the pitch lines.

While the above latent vectors include information on teammates' movements, they do not consider the opponents' behaviors. Hence, we apply an ST-Encoder to the entire player features to get FPE embeddings consulting the overall game context, i.e.,
\begin{equation}
    (\tilde{\mathbf{z}}_{g,t}^{p_1}, \ldots, \tilde{\mathbf{z}}_{g,t}^{p_{2n+4}}) = \text{ST-Encoder} (\mathbf{x}_t^{p_1}, \ldots, \mathbf{x}_t^{p_{2n+4}})
    \label{eq:fpe}
\end{equation}

Moreover, we input all the player features to a Set Transformer including encoder and decoder parts to get the FPI embedding $\tilde{\mathbf{z}}_{g,t}$ for the game context, i.e.,
\begin{equation}
    \tilde{\mathbf{z}}_{g,t} = \text{SetTransformer} (\mathbf{x}_t^{p_1}, \ldots, \mathbf{x}_t^{p_{2n+4}})
    \label{eq:fpi}
\end{equation}

Then, player-wise Bi-LSTMs with shared weights update the joint hidden states $\mathbf{h}_{g,t}^p = (\mathbf{h}_{g,t}^{p,f}, \mathbf{h}_{g,t}^{p,b})$ for each component $p \in \tilde{P} = P_1 \cup P_2 \cup P_0$ using the input features $\{ \mathbf{x}_{t}^p \}$, the PPE embeddings $\{ \mathbf{z}_{g,t}^p \}$, the FPE embeddings $\{ \tilde{\mathbf{z}}_{g,t}^p \}$, and the FPI embeddings $\{ \tilde{\mathbf{z}}_{g,t} \}$ to predict the player possession probabilities $\hat{g}_t^p$ as follows:
\begin{IEEEeqnarray}{rCl}
    \mathbf{h}_{g,t}^{p,f}  & = & \text{LSTM}^f (\mathbf{x}_t^p, \mathbf{z}_{g,t}^p, \tilde{\mathbf{z}}_{g,t}^p, \tilde{\mathbf{z}}_{g,t}; \mathbf{h}_{g,t-1}^{p,f}) \\
    \mathbf{h}_{g,t}^{p,b}  & = & \text{LSTM}^b (\mathbf{x}_t^p, \mathbf{z}_{g,t}^p, \tilde{\mathbf{z}}_{g,t}^p, \tilde{\mathbf{z}}_{g,t}; \mathbf{h}_{g,t+1}^{p,b}) \\
    \hat{g}_t^p & = & \text{FC} (\mathbf{h}_{g,t}^{p})
\end{IEEEeqnarray}

\subsubsection{Ball Trajectory Regressor} \label{se:model_btr}
For the final trajectory prediction, the model produces a partially permutation-invariant embedding (PPI) by deploying a Set Transformer for each team. Here, the hidden states $\mathbf{h}_{g,t}^p$ and the possession probabilities $\hat{g}_t^p$ resulting from PPC are concatenated with each player’s input features $\mathbf{x}_t^p$ to pass through the layer. That is,
\begin{IEEEeqnarray}{rCl}
    \mathbf{z}_t^{P_1}      & = & \text{SetTransformer} (\tilde{\mathbf{x}}_t^{p_1}, \ldots, \tilde{\mathbf{x}}_t^{p_n}) \\
    \mathbf{z}_t^{P_2}      & = & \text{SetTransformer} (\tilde{\mathbf{x}}_t^{p_{n+1}}, \ldots, \tilde{\mathbf{x}}_t^{p_{2n}}) \\
    \mathbf{z}_t^{P_0}      & = & \text{FC} (\tilde{\mathbf{x}}_t^{p_{2n+1}}, \ldots, \tilde{\mathbf{x}}_t^{p_{2n+4}}) \\
    \tilde{\mathbf{z}}_t    & = & \text{FC} (\mathbf{z}_t^{P_1}, \mathbf{z}_t^{P_2}, \mathbf{z}_t^{P_0})
\end{IEEEeqnarray}
where $\tilde{\mathbf{x}}_t^p = (\mathbf{x}_t^p, \mathbf{h}_{g,t}^p, \hat{g}_t^p)$ for $p \in \tilde{P}$.

Finally, the ball-possession-aware context embedding $\tilde{\mathbf{z}}_t$ goes into a Bi-LSTM that returns a hidden state $\mathbf{h}_t = (\mathbf{h}_t^f, \mathbf{h}_t^b)$, which is eventually converted into the predicted ball location $\hat{\mathbf{y}}_t$ by a fully-connected layer, i.e.,
\begin{IEEEeqnarray}{rCl}
    \mathbf{h}_t^f & = & \text{LSTM}^f (\tilde{\mathbf{z}}_t; \mathbf{h}_{t-1}^f) \\
    \mathbf{h}_t^b & = & \text{LSTM}^b (\tilde{\mathbf{z}}_t; \mathbf{h}_{t+1}^b) \\
    \hat{\mathbf{y}}_t  & = & \text{FC} (\mathbf{h}_t)
\end{IEEEeqnarray}

\subsection{Loss Function} \label{se:loss_function}

Our hierarchical models are trained using the loss function consisting of three parts. The first term is the \emph{mean squared error (MSE) loss} commonly used in regression tasks.
\begin{equation}
    \mathcal{L}^{\text{MSE}}(\hat{\mathbf{y}}_{1:T}, \mathbf{y}_{1:T}) = -\frac{1}{T} \sum_{t=1}^T \| \hat{\mathbf{y}}_t - \mathbf{y}_t \|_2^2
\end{equation}

\begin{figure}[htb]
    \centering
    \includegraphics[width=.47\textwidth]{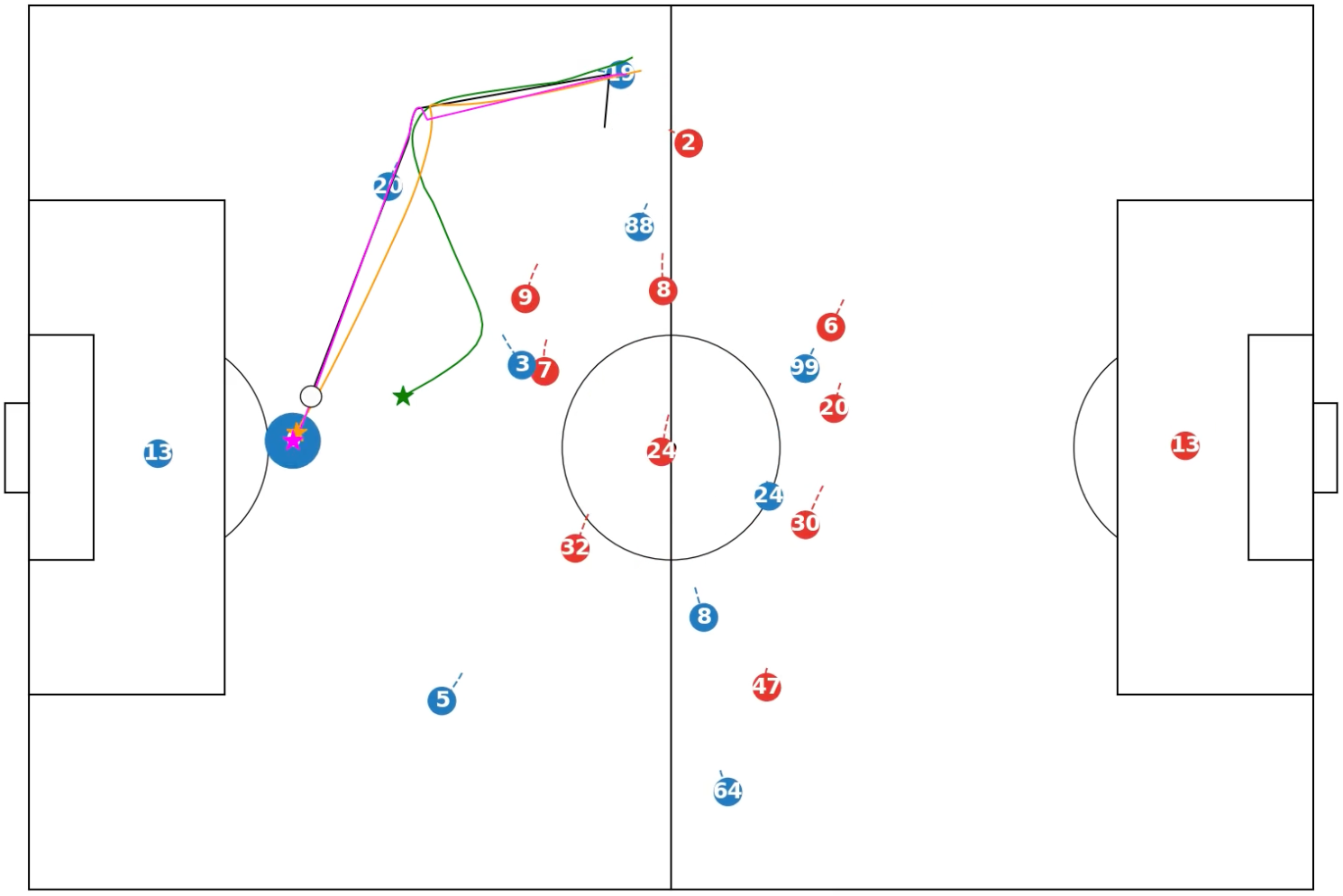}
    \caption{A match snapshot in the test dataset with the true and predicted ball trajectories. The white circle with a black tail is the ground truth, and the green, orange, and pink trajectories are predictions resulting from H-LSTM, H-LSTM-RL, and H-LSTM-RL-PP, respectively. (See the naming rule of models in Section~\ref{se:models}.) The red and blue circles are the players from either team and their sizes indicate the estimated ball possession probabilities.}
    \label{fig:snapshot_ball_pred}
\end{figure}

While the MSE forces the model to predict trajectories close to the true ones, it does not guarantee that the predicted trajectories are physically realistic. Accordingly, the model trained only using the MSE tends to make unrealistic trajectories with changes of direction even when there is no player to control the ball as the green curve in Fig.~\ref{fig:snapshot_ball_pred}. Therefore, we design the \emph{reality loss} as the second term under the assumption that the ball does not drastically change direction if there is no player close enough to the ball.

To elaborate, we first calculate the course angle of the predicted ball trajectory $\hat{\mathbf{y}}_{1:T}$ by
\begin{equation*}
    \theta_t = \arccos \left( \frac{\mathbf{v}_t \cdot \mathbf{v}_{t+1}}{\|\mathbf{v}_t\| \|\mathbf{v}_{t+1}\|} \right)
\end{equation*}
where $\mathbf{v}_t = \hat{\mathbf{y}}_{t} - \hat{\mathbf{y}}_{t-1}$ is the velocity of the ball. Also, we denote the distance between the ball and the nearest player as
\begin{equation*}
    d_t = \min_{p \in P} \| \hat{\mathbf{y}}_{t} - \mathbf{x}_t^p \|
\end{equation*}
Using $\theta_t$ and $d_t$, the reality loss is defined as
\begin{equation}
    \mathcal{L}^{\text{Real}}(\hat{\mathbf{y}}_{1:T}; X_{1:T}^{P}) = -\frac{1}{T-2} \sum_{t=2}^{T-1} \tanh(\theta_t) \cdot d_t \label{eq:real_loss}
\end{equation}

Intuitively, the reality loss increases in the situation that the ball changes its heading direction when there is no player close to it. One can train the model to return more realistic trajectories (i.e., tending to change the heading direction only near a player) by adding the reality loss term than only using the MSE loss.

An accurate intermediate prediction of player ball possession leads to an accurate prediction of the final ball trajectory. Thus, we add the \emph{cross-entropy loss}
\begin{equation}
    \mathcal{L}^{\text{CE}}(\hat{\mathbf{g}}_{1:T}, \mathbf{g}_{1:T}) = -\frac{1}{T} \sum_{t=1}^T \sum_{k=1}^K g_{t}^k \log \hat{g}_t^k
\end{equation}
for predicting ball possession as the last term of the loss function.

In summary, the total loss function is defined as
\begin{equation}
    \mathcal{L} = \mathcal{L}^{\text{MSE}} + \lambda^{\text{Real}} \mathcal{L}^{\text{Real}} + \lambda^{\text{CE}} \mathcal{L}^{\text{CE}}
\end{equation}
where $\lambda^{\text{Real}}, \lambda^{\text{CE}} \ge 0$ are weights for controlling the impact of the corresponding loss terms on the total loss, respectively.

\subsection{Rule-Based Postprocessing} \label{se:postprocess}

Although the estimated trajectories by our model are roughly similar to the true trajectories in terms of position error at each time, they may not be realistic since the neural network does not strictly enforce the output to follow physical constraints. Note that the reality loss term actually helps the output not to exhibit absurd movements but often fails to limit the trajectory to be realistic as a whole. In addition, the resulting ball possession and trajectory predictions do not provide information about whether the ball is in control by a player or moving from one player to another at a given time by the definition described in Section~\ref{se:player_poss}. Hence, we execute a rule-based postprocessing algorithm to decide whether a player is carrying the ball or it is in transition from one to another, and fine-tune the predicted trajectory based on this information.

To be specific, we obtain the \emph{possession score} $s_t^p$ for each player $p$ at time $t$ by dividing the possession probabilities $\hat{g}_t^p$ by the distance $\| \hat{\mathbf{y}}_t - \mathbf{x}_t^p \|$ from the predicted ball location. Then, we can deem that the ball has not arrived at the possessor when the possession probability is high but the predicted ball is far from the player by the following rules:
\begin{itemize}
    \item If $\hat{s}_{\tau}^q = \max_{\tilde{P}} \hat{s}_{\tau}^p > 0.5$, the player $q$ touches the ball at $\tau$.
    \item If $0.2 < \hat{s}_{\tau}^q = \max_{\tilde{P}} \hat{s}_{\tau}^p \le 0.5$ and $\hat{s}_{\tau}^q$ is a local maximum of the function $\max_{\tilde{P}} \hat{s}_t^p$ of $t$, the player $q$ touches the ball at $\tau$ (observing that there is not enough time for the possession probability to increase in one-touch pass situations).
    \item Otherwise, the ball is moving from one player to another.
\end{itemize}

\begin{figure}[hbt]
    \centering
    \begin{subfigure}[t]{0.47\textwidth}
        \includegraphics[width=\textwidth]{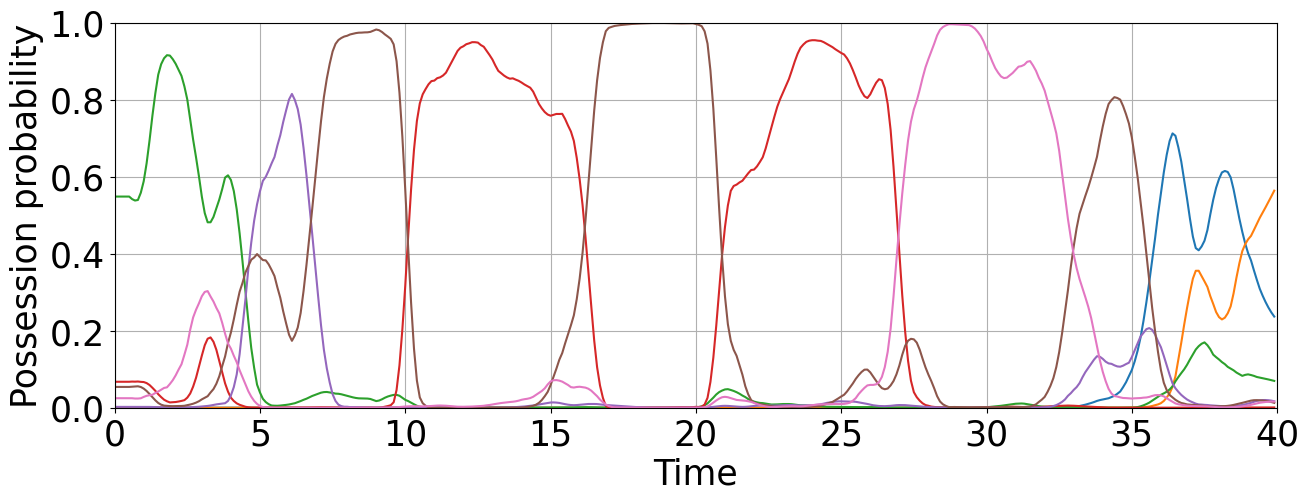}
        \caption{Possession probability plot colored by player ID.}
        \label{fig:poss_probs}
    \end{subfigure}
    \begin{subfigure}[t]{0.47\textwidth}
        \centering
        \includegraphics[width=\textwidth]{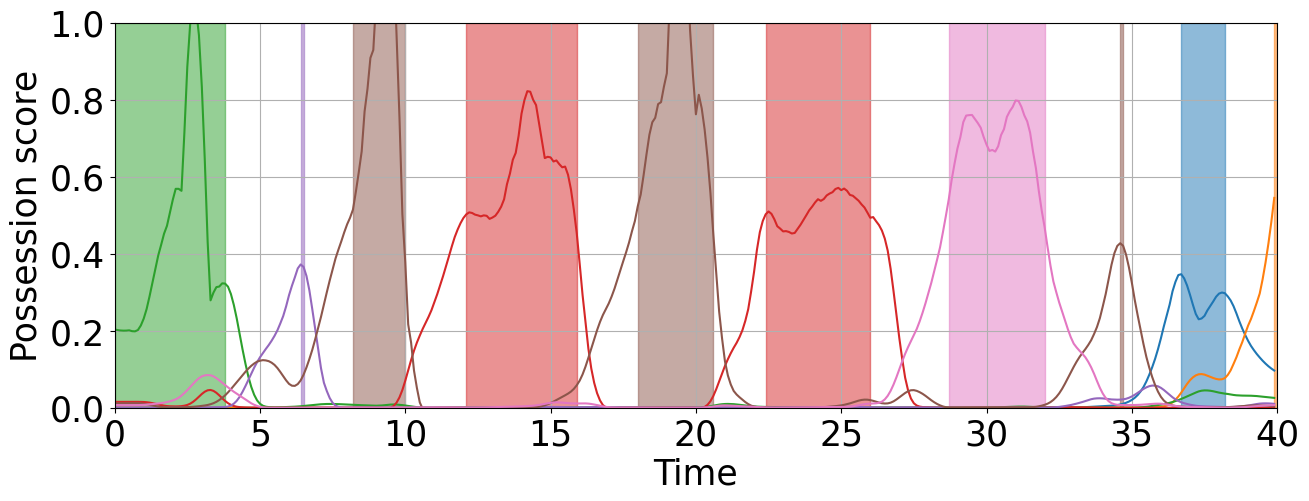}
        \caption{Possession score plot colored by player ID.}
        \label{fig:poss_scores}
    \end{subfigure}
    \caption{Possession probability and score plots in a match scene. The shaded areas in \ref{fig:poss_scores} indicate the time intervals when our rule estimates that a player is carrying the ball.}
    \label{fig:poss_plots}
\end{figure}

Figure~\ref{fig:poss_plots} shows the possession probability and score plots in a sample time interval. The possession scores in \ref{fig:poss_scores} are sharper than the probabilities in \ref{fig:poss_probs}, so we can distinguish transition intervals from touched intervals based on the score values.

After partially assigning ball-touching players by the above rule, we set the ball locations for the ``assigned'' time steps to the ball-touching players. Lastly, we reconstruct the entire trajectory by linear interpolation for the unassigned time intervals. Figure~\ref{fig:snapshot_ball_pred} shows two examples of raw predictions from either model trained with (orange) or without (green) the reality loss mentioned in Section~\ref{se:loss_function} and the postprocessed trajectory (pink) of the former model with the reality loss. One can observe that both the reality loss term and the postprocessing step contribute to the framework to return more natural trajectories in that the ball only changes its direction when it is carried or kicked by a player.

\section{Experiments} \label{se:experiments}

In this section, we implement several baseline frameworks and compare the performance with that of ours to make our design choices more compelling.

\subsection{Data Preparation} \label{se:data}

To show that our model is applicable to data from various sources, we use the mixed dataset including GPS tracking data measured by Fitogether~\cite{Fitogether} from 15 matches in K League 2020, the Korean professional soccer league, and publicly accessible optical tracking and event data\footnote{\url{https://github.com/metrica-sports/sample-data}} acquired from 3 sample matches provided by Metrica Sports~\cite{MetricaSports}. We perform the train-test split as follows:
\begin{itemize}
    \item Training data: 10 matches of Fitogether's GPS data and Metrica Sample Game 1 and 2
    \item Validation data: 2 matches of Fitogether's GPS data and the first half of Metrica Sample Game 3
    \item Test data: 3 matches of Fitogether's GPS data and the second half of Metrica Sample Game 3
\end{itemize}
For Metrica Sample Game 3, there are mismatch errors between player IDs in tracking data and event data. Thus, we have reassigned the player IDs in the event records based on the distances between the ball and the players and uploaded the corrected version to our GitHub repository\footnote{\url{https://github.com/pientist/ballradar.git}}.

\begin{table}[hbt]
  \caption{Datasets used in our experiments.}
  \label{tab:dataset}
  \centering
  \begin{tabular}{ll|cccc}
    \toprule
    {\bf Provider} & {\bf Split} & {\bf \#. Matches} & {\bf \#. Episodes} & {\bf \#. Frames} \\
    \midrule 
    Fitogether  & Train.    & 10    & 533   & 200,083 \\
                & Valid.    & 2     & 150   & 50,708 \\
                & Test      & 3     & 222   & 72,191 \\
    \midrule
    Metrica     & Train.    & 2     & 159   & 71,791 \\
                & Valid.    & 0.5   & 35    & 21,120 \\
                & Test      & 0.5   & 35    & 21,465 \\
    \bottomrule
\end{tabular}
\end{table}

One challenge is that goalkeepers (GK) are not measured by GPS trackers very often. Therefore, to predict the ball trajectory when only outfield players’ trajectories are given in real-world situations, we also train the GK trajectory prediction model using the Metrica data and use the inferred GK trajectories by applying the model to the Fitogether data. (More details about the GK trajectory prediction model are described in Appendix~\ref{se:gk_pred}.) After generating the GK trajectories, we combine the manually annotated event data with GPS-based player trajectories to reconstruct the ball trajectories to be used as the ground truth. The way of rule-based reconstruction is the same as that in Section~\ref{se:postprocess}.

Each example is a 10-second window of \SI{10}{\hertz} trajectories (i.e., 100 time steps) collected in a soccer match. The original Metrica data is \SI{25}{\hertz}, but we downsampled them to \SI{10}{\hertz} to adjust the frequency to the Fitogether data. Since there is no need to find the trajectories of balls out of play, we only take in-play situations into account and call a time interval from resuming the game to a pause an \emph{episode}. We construct the dataset by sliding a window in each episode by \SI{0.1}{\second} and sampling it at a time so that 99 time steps of a window are overlapped with the adjoining one. To avoid overfitting, we randomly flip pitches either horizontally, vertically, or in both directions. Also, we calculate the velocity, speed, and acceleration of each player and attach them to the location as input features to help the model understand the player dynamics. In summary, a training sample is a window of length 100 with 22 players having six features, the $(x,y)$ location, the $(x,y)$ velocity, the speed, and the acceleration.

\subsection{Models and Hyperparameters} \label{se:models}

The following lists are the baselines used in the experiment. They are distinguished by whether having a hierarchical architecture and the type of sequence model deployed to make predictions.
\begin{itemize}
    \item \textbf{VRNN:} A generative baseline using VRNN~\cite{Chung2015} that generates ball trajectories conditioned on player trajectories. The detailed architecture is described in Appendix~\ref{se:vrnn}.
    \item \textbf{Transformer:} A non-hierarchical framework proposed by Gongora~\cite{Gongora2021} using Transformer~\cite{Vaswani2017} for prediction.
    \item \textbf{LSTM:} A non-hierarchical framework using a Bi-LSTM.
    \item \textbf{H-Transformer:} A hierarchical framework using Transformers in both submodels.
    \item \textbf{H-LSTM:} Our hierarchical Bi-LSTM framework.
\end{itemize}
All baselines missing the prefix `H-' directly predict ball trajectories without a hierarchical architecture and have the same PPI encoder as in Section \ref{se:model_btr} to represent the game contexts. On the contrary, those with the prefix `H-' are hierarchical models that first estimate the player-level ball possession probabilities and predict the final ball trajectories conditioned on them. They employ the same context-encoding structures as in Section \ref{se:framework}, but use different sequence models (i.e., Transformer or Bi-LSTM).

To figure out the influence of reality loss, we train each hierarchical model with and without the reality loss term, respectively. The suffix `-RL' in Table~\ref{tab:main} indicates that the model is trained with $\lambda^{\text{Real}} = 1$, and the model without `-RL' takes $\lambda^{\text{Real}} = 0$. Also, we compare the prediction performance of the model before and after the rule-based postprocessing, differentiating them by attaching the suffix `-PP' to the latter.

We train each model using the Adam optimizer~\cite{Kingma2015} with an initial learning rate of 0.0005. For hierarchical models, the context embeddings $\{ \mathbf{z}_{g,t}^{p} \}_{p \in \tilde{P}}$, $\{ \tilde{\mathbf{z}}_{g,t}^{p} \}_{p \in \tilde{P}}$, and $\tilde{\mathbf{z}}_{g,t}$ in PPC have the dimension 16, while the dimension of $\{ \mathbf{z}_t^{P_k} \}_{k=0}^2$ and $\tilde{\mathbf{z}}_t$ in BTR is 128. We take $\lambda^{\text{CE}} = 20$ to match the scales of the CE and MSE losses. For LSTM frameworks, every Bi-LSTM employs two layers of 256-dimensional hidden states ($\{ \mathbf{h}_{g,t}^p \}_{p \in \tilde{P}}$ or $\mathbf{h}_t$) with dropout probability 0.2. For Transformer models, each Transformer has 4 heads and takes 256-dimensional inputs.

\subsection{Evaluation Metrics} \label{se:metrics}

In this section, we introduce the evaluation metrics for the model performance. To evaluate the prediction performance of ball trajectory, we adopt the following two metrics:
\begin{itemize}
    \item \textbf{Position error (PE):} Mean distance in meters between the predicted and true ball locations, i.e., $\frac{1}{T} \sum_{t=1}^T \| \hat{\mathbf{y}}_t - \mathbf{y}_t \|_2$.
    \item \textbf{Reality loss (RL):} Same as the reality loss introduced in Eq.~\ref{eq:real_loss} to evaluate the influence of the reality loss term and the rule-based postprocessing.
\end{itemize}
In addition, we separately assess the prediction performance of ball possession since it affects many other tasks such as postprocessing and pass annotation.
\begin{itemize}
    \item \textbf{Player-level possession accuracy (PPA):} Prediction accuracy of ball-possessing players from the player possession probabilities that the model produces, i.e., the proportion of $t$ such that
    \begin{equation*}
        \arg\max_{p \in \tilde{P}} \hat{g}_t^p = q_t
    \end{equation*}
    where $q_t \in \tilde{P}$ denotes the true player having the ball at $t$.
    \item \textbf{Team-level possession accuracy (TPA):} Prediction accuracy of attacking teams from the team possession probabilities obtained by summing up the player possession probabilities per team, i.e., the proportion of $t$ such that
    \begin{equation*}
        \arg\max_{p \in \tilde{P}} \hat{g}_t^p \in Q_t
    \end{equation*}
    where $Q_t \in \{ P_1, P_2, P_0 \}$ denotes the team that the true ball-possessing player $q_t \in \tilde{P}$ at $t$ belongs to.
\end{itemize}

\subsection{Results and Discussion}

Table~\ref{tab:main} shows the main results of our experiments, and we have found several observations from them as follows:
\begin{itemize}
    \item \textbf{Performance of the generative baseline (VRNN):} Note that when the latent vector $\mathbf{z}_t$ of VRNN is sampled from the encoder $q_{\phi}$, the mean position error is less than \SI{1}{m}. Nevertheless, the prediction using $\mathbf{z}_t$ sampled from the prior $p_{\theta}$ is far worse than LSTM or Transformer baselines. The reason we think is that unlike the previous studies for trajectory prediction or imputation in team sports~\cite{Felsen2018, Omidshafiei2022, Yeh2019, Zhan2019}, the prior in our problem cannot leverage any fragmentary trajectory of the target. This hinders the model from reducing the KL divergence between $p_{\theta}$ and $q_{\phi}$.
    \item \textbf{Performance of the Transformer baselines:} The Transformer baselines show lower performance than those of their LSTM counterparts. We think the use of attentions instead of recurrence backfires in our problem. Transformers learn which time steps to focus on by multi-head attentions, showing great effects in many problems by resolving the information bottleneck imposed on the last hidden state of the encoder. However, paradoxically, the latest player locations are the most important context to predict the ball location, so the inclination of RNN models to focus on the latest time step is rather helpful in our case. Also, the recurrence seems to reduce the reality loss since it more strongly connects the adjacent time steps than the attention does.
    \item \textbf{Effects of adopting the hierarchical architecture:} Hierarchical models (H-Transformer and H-LSTM) exhibit better performance than their non-hierarchical counterparts (Transformer and LSTM). This implies that the introduction of the intermediate target variable (player-level ball possession) promotes more accurate ball trajectory prediction.
    \item \textbf{Effects of the reality loss:} Training a model with the reality loss term is shown to have generalization power in that it reduces the RL of predicted ball trajectories of the test data. In addition, it slightly improves essential performance in terms of PE and PPA. 
    \item \textbf{Effects of rule-based postprocessing:} The postprocessing step seems to sacrifice position accuracy to retain the naturalness of output trajectories. One might think that it is better not to perform postprocessing because it increases PE. However, it enables event detection as described in Section~\ref{se:pass_annot} by separating a player's ball-possessing interval into a ``transition'' period and a ``controlled'' period. Thus, ultimately it is a necessary step since analyzing ball-related events is one of the main purposes of ball data acquisition.
\end{itemize}

\begin{table}[hbt]
  \caption{Performance of baseline models for ball trajectory and player/team-level ball possession prediction.}
  \label{tab:main}
  \centering
  \setlength{\tabcolsep}{3.5pt}
  \begin{tabular}{l|cccc}
    \toprule
    {\bf Model} & {\bf PE} &  {\bf RL} & {\bf PPA} & {\bf TPA} \\
    \midrule
    VRNN        & 12.3721 & 0.7206 & - & - \\
    Transformer & 7.4727  & 1.2194 & - & - \\
    LSTM        & 5.4667  & 0.3392 & - & - \\
    \midrule
    H-Transformer       & 4.9136 & 0.7987 & \SI{48.91}{\%} & \SI{77.91}{\%} \\
    H-Transformer-RL    & 4.6933 & 0.5990 & \SI{49.52}{\%} & \SI{78.01}{\%} \\
    H-Transformer-PP    & 5.4075 & 0.0021 & \SI{52.01}{\%} & \SI{78.78}{\%} \\
    H-Transformer-RL-PP & 5.4093 & 0.0021 & \SI{52.18}{\%} & \SI{78.66}{\%} \\
    \midrule
    H-LSTM       & 3.6886 & 0.1881 & \SI{64.01}{\%} & \SI{85.26}{\%} \\
    H-LSTM-RL    & 3.6561 & 0.1391 & \SI{64.70}{\%} & \SI{85.85}{\%} \\
    H-LSTM-PP    & 4.1285 & 0.0016 & \SI{63.84}{\%} & \SI{85.34}{\%} \\
    H-LSTM-RL-PP & 4.0719 & 0.0017 & \SI{64.32}{\%} & \SI{85.34}{\%} \\
    \bottomrule
\end{tabular}
\end{table}

\section{Practical Applications} \label{se:apps}

In this section, we provide a list of potential use cases of our framework, including missing trajectory imputation in the situation of video-based ball detection (Section \ref{se:imputation}), semi-automated pass annotation based on the estimated ball possession (Section \ref{se:pass_annot}), automated zoom-in for match broadcasting based on the estimated ball locations (Section \ref{se:auto_zoomin}), and separating running performance metrics into attacking and defending phases using the aggregated team possession probabilities (Section \ref{se:rp_metrics}).

For Section \ref{se:imputation}, \ref{se:pass_annot}, and \ref{se:rp_metrics}, we use three test matches of Fitogether data as described in Section~\ref{se:data}. On the other hand, we only use one match among the three in Section \ref{se:auto_zoomin} since it is the only match in the test dataset where we have GPS data, event data, and a full-pitch video recorded by fixed cameras.

\subsection{Ball Trajectory Imputation} \label{se:imputation}

A major challenge of computer vision-based ball tracking is that the estimated ball trajectory is often inaccurate, especially when the ball moves very fast or is occluded by other objects such as players. A plausible approach to this problem is to find ``reliable'' fragments of ball trajectories obtained by object detection and perform imputation leveraging our framework. Namely, we first choose frames where the ball is clearly observable without any occlusion and in a relatively stationary situation. Then, we interpolate the ball locations for the remaining ``unreliable'' frames conditioned on the players’ trajectories and the fragmentary ball trajectories. More specifically, to adapt our framework to this scenario, we randomly mask the true ball trajectory by a certain probability and train our H-LSTM-RL to estimate the ball locations for the missing frames with the masked trajectories.

\begin{table}[t]
  \caption{Trajectory imputation performance of the proposed model before (H-LSTM-RL) and after (H-LSTM-RL-PP) the postprocessing with varying masking probabilities.}
  \label{tab:imputation}
  \centering
  \begin{tabular}{cc|cccc}
    \toprule
    \textbf{Step} & \textbf{Masking} & \textbf{PE} & \textbf{RL} & \textbf{PPA} & \textbf{TPA} \\
    \midrule
    Before PP   & \SI{100}{\%} & 5.3536 & 0.3573 & \SI{61.91}{\%} & \SI{83.96}{\%} \\
    &  \SI{95}{\%} & 3.0018 & 0.3911 & \SI{79.51}{\%} & \SI{91.54}{\%} \\ 
    &  \SI{90}{\%} & 2.0939 & 0.4220 & \SI{87.11}{\%} & \SI{95.04}{\%} \\
    &  \SI{80}{\%} & 1.3052 & 0.5059 & \SI{93.13}{\%} & \SI{97.41}{\%} \\
    \midrule
    After PP    & \SI{100}{\%} & 5.1440 & 0.0031 & \SI{57.29}{\%} & \SI{81.06}{\%} \\
    &  \SI{95}{\%} & 3.7990 & 0.0056 & \SI{78.71}{\%} & \SI{90.73}{\%} \\ 
    &  \SI{90}{\%} & 2.5913 & 0.0046 & \SI{85.92}{\%} & \SI{94.50}{\%} \\
    &  \SI{80}{\%} & 1.4137 & 0.0038 & \SI{91.25}{\%} & \SI{96.67}{\%} \\
    \bottomrule
\end{tabular}
\end{table}

During training, the model is provided with partial trajectories with \SI{80}{\%} of the values masked for half of the batches, while it does not refer to any target trajectories for the other half. Then, we measure the prediction performance of the trained model for the test data given partial ball trajectories with varying masking probabilities (\SI{100}{\%}, \SI{95}{\%}, \SI{90}{\%}, and \SI{80}{\%}).

According to the result in Table \ref{tab:imputation}, the prediction performance seems to dramatically improve when only \SI{10}{\%} of ground-truth is given. When we provide \SI{20}{\%} of targets to the model, PE reduces to less than \SI{1.5}{m} and PPA rises to more than \SI{90}{\%}. This improvement implies that our method can be successfully employed to impute or revise the missing or unreliable results of other ball detection or tracking frameworks by leveraging multi-agent game contexts.

\subsection{Semi-automated Pass Annotation} \label{se:pass_annot}

One of the ultimate goals of ball tracking in soccer, a representative team sports, is to detect and analyze event data that occurred during the match. Soccer event data is a record of on-the-ball actions such as passes, interceptions, dribbles, and shots that occurred during matches, originally collected by human annotators~\cite{Decroos2020}. It is actively used from aggregating match statistics~\cite{FIFAWC2022} to various tasks including performance evaluation~\cite{Bransen2020,Pappalardo2019,Decroos2019,Luo2020}, playing style representation~\cite{Decroos2019a,Decroos2020a,Cho2022,Clijmans2022}, tactical analysis~\cite{Wang2015,Decroos2018}, and so on. However, since 2,000 events occur in a match, data collection requires a lot of manual work.

Several studies have tried to automatically detect events in soccer matches. Codina et al.~\cite{Codina2022} introduced a rule-based framework for event detection from player and ball tracking data. Sorano et al.~\cite{Sorano2020} studied the problem of detecting passes from soccer videos by utilizing a CNN-based object detection engine (YOLOv3) and Bi-LSTM. Fassmeyer et al.~\cite{Fassmeyer2021} proposed a method that detects a wider range of events such as corner kicks, crosses, and counterattacks using a Variational Autoencoder (VAE) and Support Vector Machine (SVM). Note that all of these approaches rely on the ball tracking information from video data~\cite{Sorano2020,Codina2022} or manual annotation~\cite{Fassmeyer2021}.

Leveraging our framework, we also perform the event detection task using the predicted ball trajectory instead of the true ball trajectory. For simplicity, we only detect passes, the most frequent and fundamental event type in soccer matches. From the ball touch information predicted in Section~\ref{se:postprocess}, we define passes and their successes by a naive rule as follows:
\begin{definition}[Pass]
    If a player $p$ touches the ball at $t_0$, another player $q$ touches the ball at $t_1 > t_0$, and the ball is in transition in $(t_0, t_1)$, then we define there is a \emph{pass} from $p$ to $q$ in $(t_0, t_1)$.
\end{definition}

We calculate pass detection accuracy regarding that a true pass from $p$ to $q$ in $(t_0, t_1)$ is correctly detected if there is a detected pass from $p$ to $q$ starting after $t_0 - \SI{2}{\second}$ and ending before $t_1 + \SI{2}{\second}$. Also, we evaluate the passer and receiver detection accuracies by counting the passes that at least the passer or receiver is correct. Since the number of passes is a common statistic in soccer match analysis, we aggregate the numbers of passes and receives per player and compare them to the true values, too.

\begin{table}[htb]
  \caption{Pass annotation accuracies, including the F1 scores for pass, passer, and receiver detection tasks and the R2 scores for the numbers of passes and receives per player.}
  \label{tab:pass_annot}
  \centering
  \setlength{\tabcolsep}{3pt}
  \begin{tabular}{c|ccccc}
    \toprule
   {\bf Masking} & {\bf Pass} & {\bf Passer} & {\bf Receiver} & {\bf \#. Passes} & {\bf \#. Receives}\\ \midrule
    \SI{100}{\%} & 0.3877 & 0.6039 & 0.5707 & 0.7935 & 0.8150 \\
    \SI{95}{\%}  & 0.5972 & 0.7265 & 0.7199 & 0.9031 & 0.9493 \\ 
    \SI{90}{\%}  & 0.7428 & 0.8315 & 0.8208 & 0.9347 & 0.9717 \\
    \SI{80}{\%}  & 0.8693 & 0.9149 & 0.8991 & 0.9331 & 0.9746 \\
  \bottomrule
\end{tabular}
\end{table}

As a result, Table~\ref{tab:pass_annot} demonstrates that the pass detection accuracy is around \SI{87}{\%} in terms of F1 score when only \SI{20}{\%} of ground truth is given to the model. Also, the accuracy of the statistics such as the numbers of passes or receives is quite high even when the model can leverage \SI{5}{\%} of true trajectories. This implies the potential use of our framework for building a semi-automated event annotation system that the model first suggests ``candidate'' passes and human annotators only correct partial prediction errors. Considering that collecting event and ball-tracking data requires a substantial cost, this AI-assisted annotation system would also contribute a lot to data completion in the sports analytics industry.

\subsection{Automated Zoom-in for Match Broadcasting} \label{se:auto_zoomin}

\begin{figure}[tbh]
    \centering
    \includegraphics[width=0.20\textwidth]{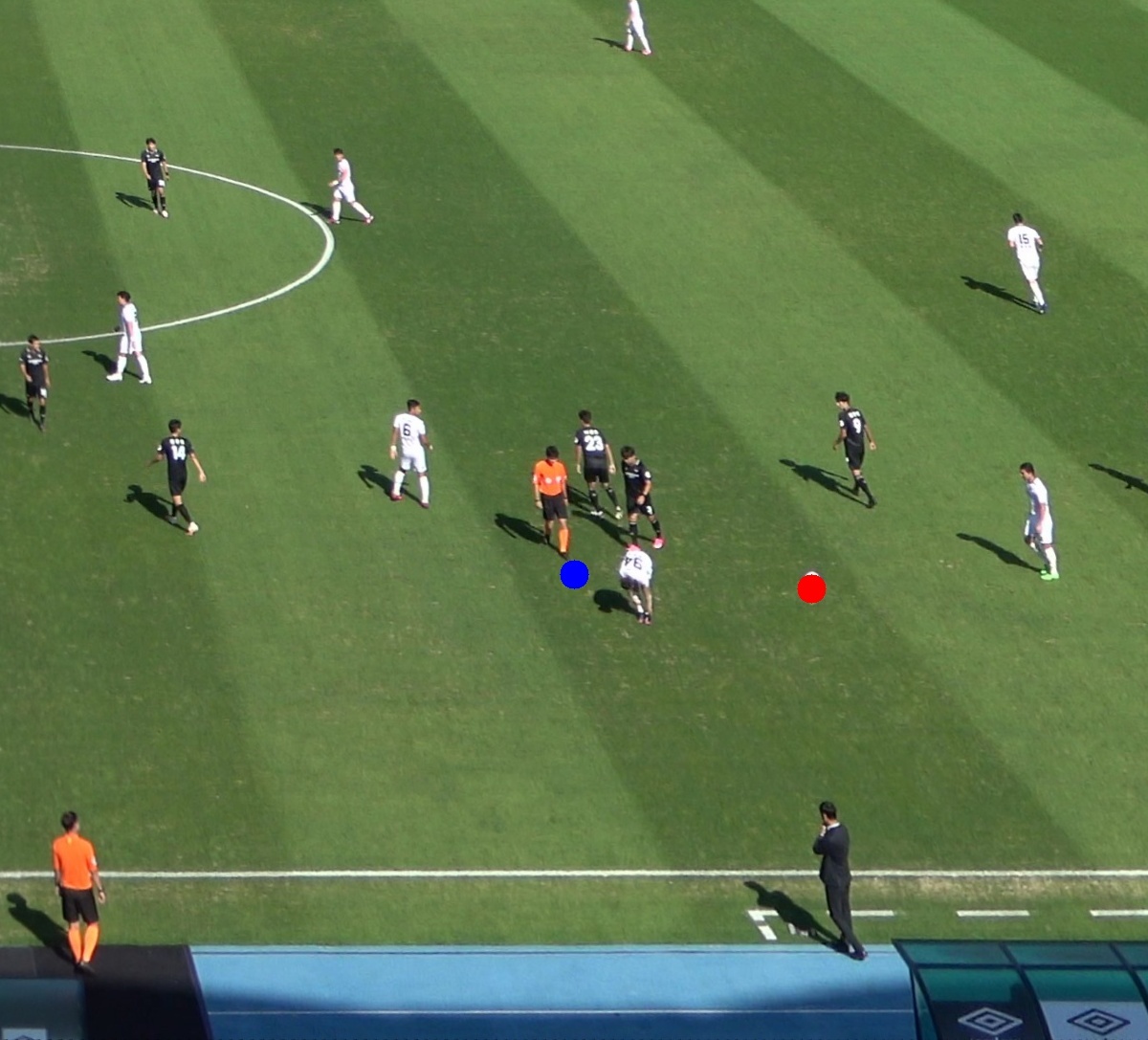}
    \includegraphics[width=0.27\textwidth]{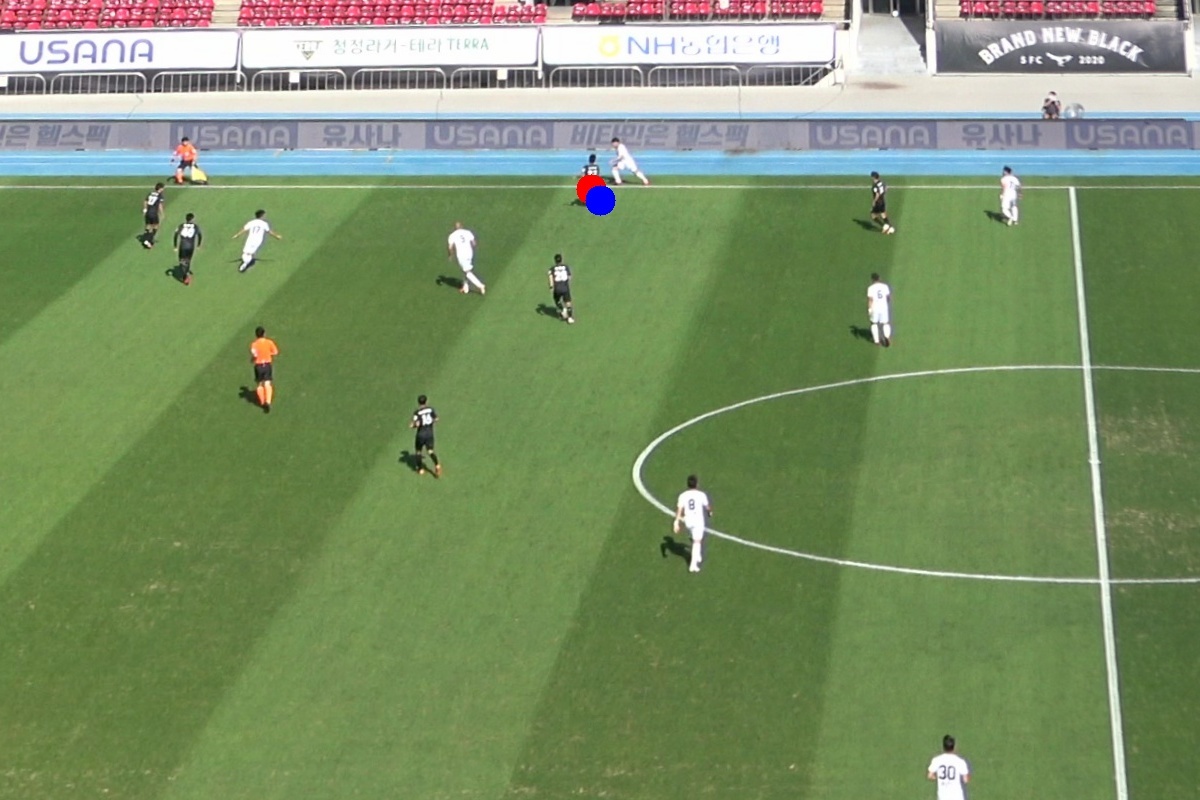}
    \caption{Example snapshots of a video resulting from recording a match by a fixed camera and automatically zooming in on the panoramic video using the estimated ball locations.}
    \label{fig:roi_examples}
\end{figure}

Automatic sports broadcasting inevitably involves tracking the ball or players to zoom in on the region of interest from the whole pitch~\cite{Yu2004,Chen2008,Pan2021}.
Figure~\ref{fig:roi_examples} demonstrates two snapshots obtained by setting the estimated ball position by our model to the center of the snapshots. Blue circles indicate the centers of snapshots while red circles mean the true ball locations in the video. As we can see from the snapshots, our model locates the ball quite well without using any result of ball detection.

\begin{table}[tbh]
    \caption{Accuracies of ROI bounding boxes of varying sizes centered on the estimated ball locations.}
    \label{tab:roi_acc}
    \setlength{\tabcolsep}{3pt}
    \centering
    \begin{tabular}{l|cccccc}
    \toprule
    {\bf Bounding box size} & 100 & 200 & 300 & 500 & 600 \\
    \midrule
    {\bf ROI accuracy} & \SI{36.37}{\%} & \SI{66.11}{\%} & \SI{79.97}{\%} & \SI{92.82}{\%} & \SI{96.25}{\%} \\
    \bottomrule
    \end{tabular}
\end{table}

Table~\ref{tab:roi_acc} shows the accuracy of the estimated ball locations in terms of region-of-interest (ROI) bounding box obtained by setting the center of a bounding box to be the estimated ball location at that time. We calculate the ``ROI accuracy'' as the proportion of frames where the estimated ROI contains the true ball. We can observe that the ROI accuracy reaches about \SI{80}{\%} when the size of the bounding box is $300 \times 300$, which is the standard resolution of input images in recent object detection networks. Also, when we increase the bounding box size to $600 \times 600$, the ROI accuracy reaches \SI{96}{\%}. This implies that we may automatically generate soccer match broadcasting videos following the estimated ball trajectory instead of the real ball trajectory obtained by relying on sophisticated ball tracking algorithms, as the resolution of the broadcasting videos is usually much higher than $600 \times 600$.

\subsection{Approximating Possession-wise Running Performance Metrics} \label{se:rp_metrics}
The major use of GPS tracking data is to monitor players’ physiological demands. Running performance (RP) metrics such as total distance covered, distance covered by speed zone, and the number of sprints find general acceptance in the sports science domain as indicators for players’ workload~\cite{Cummins2013}. Moreover, several studies~\cite{Hoppe2015,Rampinini2009,Souza2020} observed that RP with ball possession has a greater influence on the team’s success than that without ball possession. Accordingly, there is a growing interest in separately calculating RP metrics in offensive and defensive phases~\cite{Modric2021}.

\begin{figure}[htb]
    \centering
    \includegraphics[width=.48\textwidth]{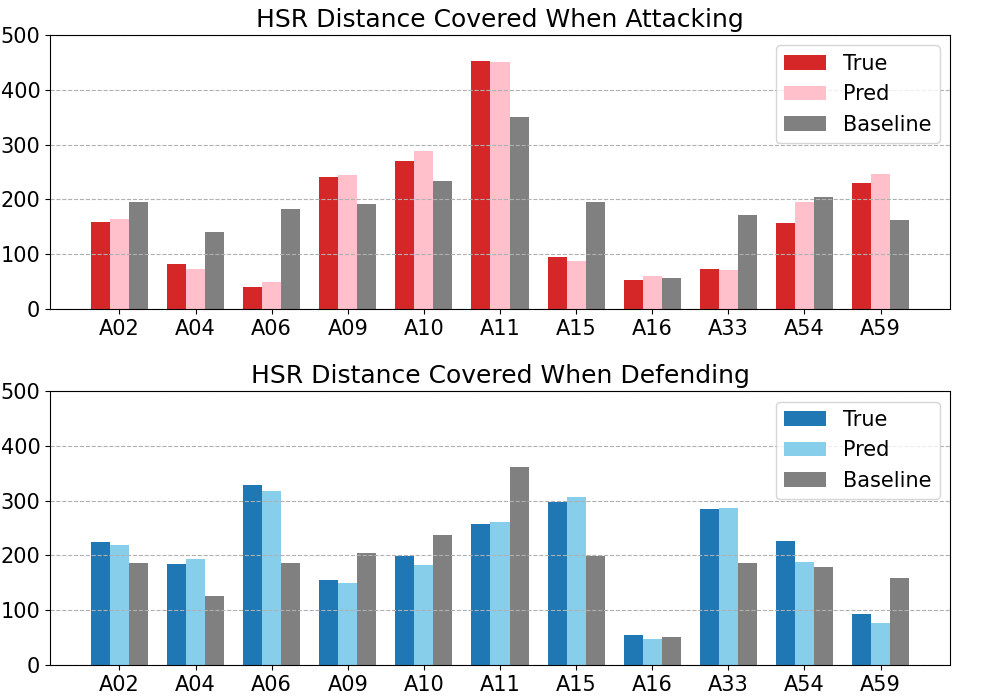}
    \caption{The estimated and true HSR distance values either when attacking or defending in a test match.}
    \label{fig:rp_metrics}
\end{figure}

In this section, we demonstrate that accurately predicted team possession can also be a useful by-product of our framework by approximating the RP metrics in attacking and defending situations. Here we estimate that the team of the predicted ball possessor at each time is attacking at that time. We calculate the total distance and high-intensity running (running with speed > $\SI{20}{\kilo\meter\per\hour}$) distance covered per player for situations that our model predicts as offensive or defensive, respectively. Also, we randomly assign a team possession label per time as a baseline. Then, we compare the offensive and defensive RP metrics resulting from each method with the ground truth by calculating the maximum absolute percentage errors (MAPE). (See Figure~\ref{fig:rp_metrics} visualizing the estimated and true HSR distance values for each of the attacking and defending phases in a test match.)

Table~\ref{tab:rp_metrics} shows that our model provides highly accurate total distances with MAPE of only about 0.035 and HSR distances with MAPE much less than the random baseline. A notable point is that the prediction accuracy for possession-wise RP metrics except for the HSR distance in attacking situations is even higher than that of raw team possession. This is because false offenses and false defenses supplement each other when aggregating the metrics and cancel out the errors.

\begin{table}[htb]
  \caption{Maximum absolute percentage errors (MAPE) for the HSR distance values estimated by our method and the random possession assignment as a baseline.}
  \label{tab:rp_metrics}
  \setlength{\tabcolsep}{4pt}
  \begin{tabular}{l|cc|cc}
    \toprule
    & \multicolumn{2}{c|}{\bf Total distance} & \multicolumn{2}{c}{\bf HSR distance} \\
    {\bf Method} & {\bf Attacking} & {\bf Defending} & {\bf Attacking} & {\bf Defending} \\
    \midrule
    Random  & 0.0881 & 0.0771 & 0.5140 & 0.2091 \\
    Ours    & 0.0360 & 0.0340 & 0.2058 & 0.1251 \\
    \bottomrule
  \end{tabular}
\end{table}

\section{Conclusion}

We address the problem of ball trajectory inference using the player context in sports matches. We combine the Set Transformer to represent the permutation-invariant and equivariant nature of team sports situations and a hierarchical recurrence structure to intermediately predict the game semantics such as ball possession. Other than the previous studies for trajectory prediction in multi-agent contexts, our framework estimates the accurate trajectory even when neither partial nor past trajectories of the target are given. Moreover, we suggest practical use cases mainly related to enhancing data acquisition and automating manual works including missing trajectory imputation, event annotation, and zoom-in on broadcasting videos. We expect that our method contributes to accumulating fundamental data for sports analytics but with a much lower level of difficulty than before.

\section*{Acknowledgements}

This work was supported by the National Research Foundation of Korea (NRF) grant funded by the Korean government (MSIT) (No. RS-2023-00208094).

\bibliographystyle{abbrv}
\bibliography{ballradar}

\newpage
\appendix

\section{Description of Subsidiary Models}

In this section, we elaborate on the subsidiary models mentioned in the paper. These include the goalkeeper (GK) trajectory prediction model introduced in Section~\ref{se:data} and the generative baseline using VRNN in Section~\ref{se:models}.

\subsection{GK Trajectory Prediction Model} \label{se:gk_pred}

First, we outline the architecture of the GK trajectory prediction model. Considering that goalkeepers' behavior depends on whether the team is attacking or defending, the model has a hierarchical architecture leveraging team-level ball possession as an intermediate target. That is, it consists of two submodels, the team possession classifier (TPC) and the GK trajectory regressor (GTR).

\begin{figure}[bth]
    \centering
    \includegraphics[width=0.45\textwidth]{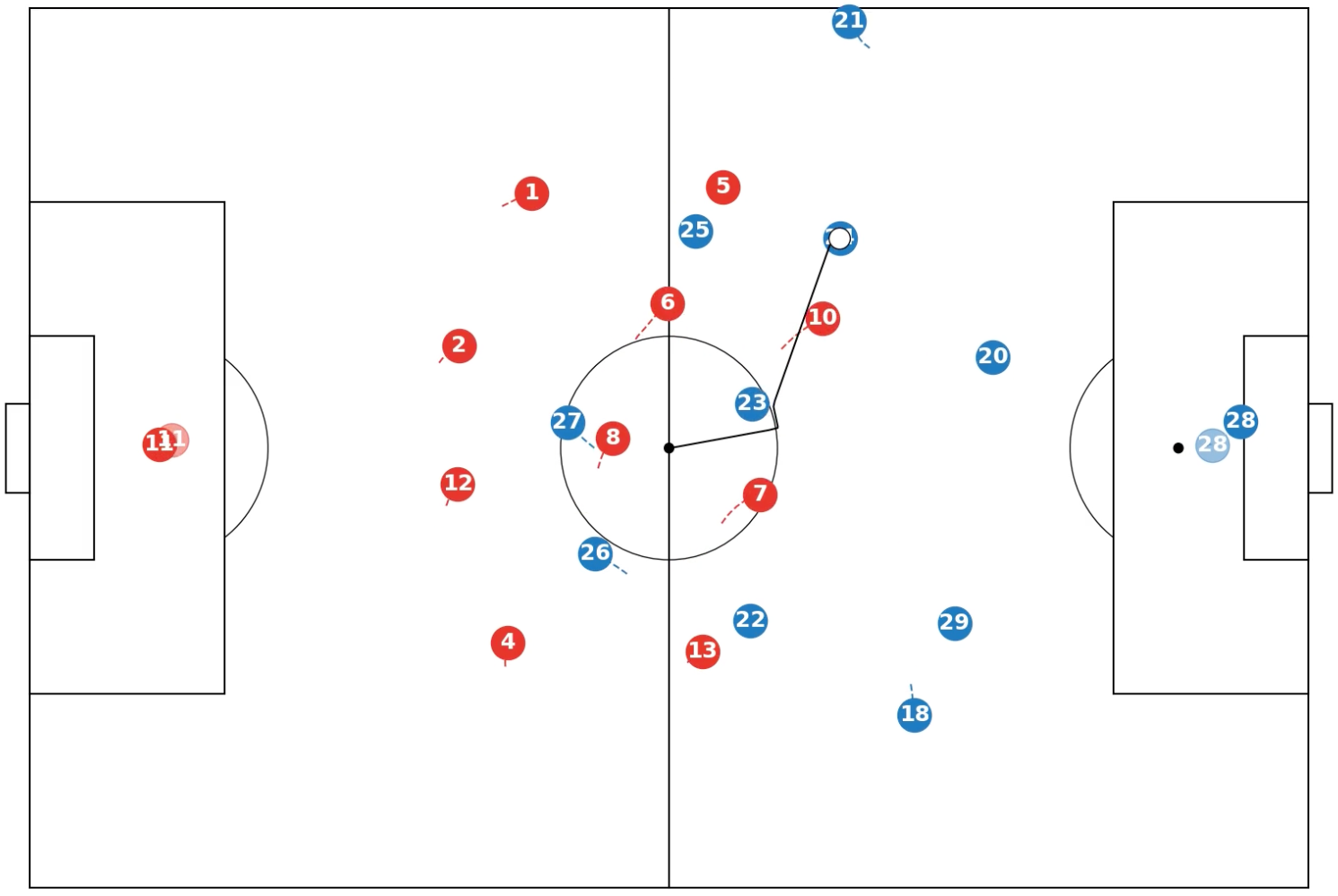}
    \caption{A snapshot of Metrica Sample Game 3, where the faint circles indicate the predicted locations of the GKs.}
    \label{fig:snapshot_gk_pred}
\end{figure}

The team-level ball possession differs from player-level ball possession in that it is partially permutation-invariant with respect to input players rather than permutation-equivariant. Thus, we deploy a single Bi-LSTM for all the players instead of constructing player-wise Bi-LSTMs. To be specific, we apply a Set Transformer for input player features in each team $P_k$, and merge them by a fully-connected layer to make a representation $\tilde{\mathbf{z}}_{t}$ of the game state.
\begin{IEEEeqnarray}{rCl}
    \mathbf{z}_{t}^{P_1}  & = & \text{SetTransformer} (\mathbf{x}_t^{p_1}, \ldots, \mathbf{x}_t^{p_n}) \label{eq:tpa_ppi_team1} \\
    \mathbf{z}_{t}^{P_2}  & = & \text{SetTransformer} (\mathbf{x}_t^{p_{n+1}}, \ldots, \mathbf{x}_t^{p_{2n}}) \label{eq:tpa_ppi_team2} \\
    \tilde{\mathbf{z}}_{t}  & = & \text{FC} (\mathbf{z}_t^{P_1}, \mathbf{z}_t^{P_2}) \label{eq:tpa_ppi_fc}
\end{IEEEeqnarray}
where $\{ p_1, \ldots, p_n \}$ and $\{ p_{n+1}, \ldots, p_{2n} \}$ are the players in team $P_1$ and $P_2$, respectively.

Then, a single Bi-LSTM updates the hidden state $\mathbf{h}_{g,t} = (\mathbf{h}_{g,t}^{f}, \mathbf{h}_{g,t}^{b})$ using the partially permutation-invariant context embedding $\tilde{\mathbf{z}}_{t}$ to predict the team possession probabilities $\hat{\mathbf{g}}_t = (\hat{g}_t^{P_1}, \hat{g}_t^{P_2})$ as follows:
\begin{IEEEeqnarray}{rCl}
    \mathbf{h}_{g,t}^f  & = & \text{LSTM}^f (\tilde{\mathbf{z}}_{t}; \mathbf{h}_{g,t-1}^f) \\
    \mathbf{h}_{g,t}^b  & = & \text{LSTM}^b (\tilde{\mathbf{z}}_{t}; \mathbf{h}_{g,t+1}^b) \\
    \hat{\mathbf{g}}_t & = & \text{FC} (\mathbf{h}_{g,t})
\end{IEEEeqnarray}

The GTR part is quite simple. It reuses the context embedding $\tilde{\mathbf{z}}_{t}$ as an input to the second Bi-LSTM along with the team probabilities $\hat{\mathbf{g}}_t$ from the previous block. The final output $\hat{\mathbf{y}}_t = (\hat{\mathbf{y}}_t^{P_1}, \hat{\mathbf{y}}_t^{P_2})$ where $\hat{\mathbf{y}}_t^{P_k} \, (k = 1,2)$ denotes the estimated location of of team $P_k$'s goalkeeper at time $t$ is obtained by passing the resulting hidden state $\mathbf{h}_t = (\mathbf{h}_t^f, \mathbf{h}_t^b)$ to a fully connected layer.

\begin{IEEEeqnarray}{rCl}
    \mathbf{h}_{t}^f  & = & \text{LSTM}^f (\tilde{\mathbf{z}}_{t}, \hat{g}_t; \mathbf{h}_{t-1}^f) \\
    \mathbf{h}_{t}^b  & = & \text{LSTM}^b (\tilde{\mathbf{z}}_{t}, \hat{g}_t; \mathbf{h}_{t+1}^b) \\
    \hat{\mathbf{y}}_t & = & \text{FC} (\mathbf{h}_{t})
\end{IEEEeqnarray}

The GK trajectory prediction model trained on the Metrica training data described in Section~\ref{se:data} shows the average position error of \SI{5.4}{m} for the Metrica test data. We apply this model to Fitogether data and use the predicted GK trajectories, together with the original outfield players' trajectories, for the ball trajectory prediction task. See Figure~\ref{fig:snapshot_gk_pred} as an example of predicted GK trajectories.

\subsection{Context-Aware VRNN as a Baseline} \label{se:vrnn}

In Section~\ref{se:models}, we implement a generative baseline and compare the prediction performance with our regression model to explain our design choices. Since major studies~\cite{Felsen2018,Yeh2019,Zhan2019,Omidshafiei2022} for player trajectory prediction or imputation have built their framework on top of Variational Recurrent Neural Network (VRNN)~\cite{Chung2015}, we also construct a generative model based on the VRNN.

\subsubsection{Difference in problem settings from the previous studies}
The previous studies deal with the problem of observing fragmentary trajectories of given agents and predicting the remaining parts of the same agents’ trajectories. On the other hand, the target agent to predict the trajectory (i.e., the ball) differs from the agents that the model refers to (i.e., players) in our problem setting. That is, our ball trajectory prediction task is clearly different from the aforementioned studies in that (1) the model can observe non-target trajectories throughout the entire time interval and (2) it cannot leverage any fragmentary trajectory of the target.

\subsubsection{Model architecture}
We basically take the autoregressive architecture of the original VRNN (as adopted for future trajectory prediction tasks~\cite{Yeh2019,Zhan2019}), but make slight changes to reflect these differences. To put it concretely, the model consists of the three deep neural networks
\begin{IEEEeqnarray}{rCl}
    f^{\text{pri}}(\mathbf{h}_{t-1};\theta) & = & [\mu^{\text{pri}}_t,\sigma^{\text{pri}}_t] \label{eq:vrnn_pri} \\
    f^{\text{enc}}(\mathbf{x}_t,\mathbf{h}_{t-1};\phi) & = & [\mu^{\text{enc}}_t,\sigma^{\text{enc}}_t] \label{eq:vrnn_enc} \\
    f^{\text{dec}}(\mathbf{z}_t,\mathbf{h}_{t-1};\psi) & = & [\mu^{\text{dec}}_t,\sigma^{\text{dec}}_t] \label{eq:vrnn_dec}
\end{IEEEeqnarray}
where $\mathbf{x}_t$ is the ball location, $\mathbf{z}_t$ is the latent state of VAE, $\mathbf{h}_t = (\mathbf{h}_t^f, \mathbf{h}_t^b)$ is the asymmetric joint hidden state of the Bi-LSTM with
\begin{IEEEeqnarray}{rCl}
    \mathbf{h}_t^f & = & \text{LSTM}^f (\tilde{\mathbf{o}}_{t}, \mathbf{x}_{t}, \mathbf{z}_{t}; \mathbf{h}_{t-1}^f) \\
    \mathbf{h}_t^b & = & \text{LSTM}^b (\tilde{\mathbf{o}}_{t}; \mathbf{h}_{t+1}^b)
\end{IEEEeqnarray}
and $\theta, \phi, \psi$ are trainable parameters.

Note that $\tilde{\mathbf{o}}_{t}$ denotes the context embedding obtained from non-target (i.e., players) trajectories $(\mathbf{x}_{t}^{p_1}, \ldots, \mathbf{x}_{t}^{p_{2n}})$, where the procedure of constructing it is the same as that of $\tilde{\mathbf{z}}_{t}$ in Appendix~\ref{se:gk_pred} (See Eq.~\ref{eq:tpa_ppi_team1}--\ref{eq:tpa_ppi_fc}). Here we use a different notation $\tilde{\mathbf{o}}_{t}$ instead of $\tilde{\mathbf{z}}_{t}$ to avoid confusion with the latent vector $\mathbf{z}_{t}$ of VRNN. Also, while the recurrence in the original VRNN operates through a unidirectional RNN (i.e., $\mathbf{h}_t = \text{RNN}(\mathbf{x}_t, \mathbf{z}_t; \mathbf{h}_{t-1})$, we add a backward RNN since the model can leverage the information of the future context $\tilde{\mathbf{o}}_{>t}$.

\subsubsection{Ball trajectory generation}
The neural network outputs in Eq.~\ref{eq:vrnn_pri}--\ref{eq:vrnn_dec} act as parameters of the following normal distributions for the latent state $\mathbf{z}_{t}$ and the target $\mathbf{x}_{t}$, respectively:
\begin{IEEEeqnarray}{rCl}
    p_{\theta}(\mathbf{z}_{t} | \tilde{\mathbf{o}}_{1:T}, \mathbf{x}_{<t}, \mathbf{z}_{<t}) & = & \mathcal{N} \left( \mathbf{z}_{t} | \mu^{\text{pri}}_t, \text{diag}(\sigma^{\text{pri}}_t)^2 \right) \\
    q_{\phi}(\mathbf{z}_{t} | \tilde{\mathbf{o}}_{1:T}, \mathbf{x}_{\le t}, \mathbf{z}_{<t}) & = & \mathcal{N} \left( \mathbf{z}_{t} | \mu^{\text{enc}}_t, \text{diag}(\sigma^{\text{enc}}_t)^2 \right) \\
    p_{\psi}(\mathbf{z}_{t} | \tilde{\mathbf{o}}_{1:T}, \mathbf{x}_{<t}, \mathbf{z}_{\le t}) & = & \mathcal{N} \left( \mathbf{x}_{t} | \mu^{\text{dec}}_t, \text{diag}(\sigma^{\text{dec}}_t)^2 \right)
\end{IEEEeqnarray}

Given the context embedding $\tilde{\mathbf{o}}_{t}$, a latent random variable $\mathbf{z}_{t}$ is sampled from the prior distribution $p_{\theta}(\mathbf{z}_{t} | \tilde{\mathbf{o}}_{1:T}, \mathbf{x}_{<t}, \mathbf{z}_{<t})$. Then, the decoder distribution $p_{\psi}(\mathbf{z}_{t} | \tilde{\mathbf{o}}_{1:T}, \mathbf{x}_{<t}, \mathbf{z}_{\le t})$ generates the ball location $\mathbf{x}_{t}$ conditioned on $\mathbf{z}_{t}$.

\section{Ablation Study} \label{se:ablation}

As well as the main experiments described in Section~\ref{se:experiments}, we carried out several ablation studies to examine the contributing factors of the proposed framework.

\subsection{Effects of Using Derivatives of Coordinates} \label{se:ablation_features}

First, we have conducted an ablation study about input features to demonstrate that the use of the first-order and second-order derivatives of the players' raw (x,y)-coordinates successfully improves our model. Table~\ref{tab:ablation_features} exhibits that our full model using six physical features (2D location, 2D velocities, speed, and acceleration) per player achieves much better performance compared to the degraded models taking four (2D location and 2D velocities) and two features (2D location only), respectively.

\begin{table}[ht!]
\caption{Performance comparison between models using different subsets of features.}
\label{tab:ablation_features}
\setlength{\tabcolsep}{2pt}
\begin{tabular}{l|cccc}
    \toprule
    {\bf Features per player}     & {\bf PE} & {\bf RL} & {\bf PPA} & {\bf TPA} \\
    \midrule
    Location                & 4.6470  & 0.1906 & \SI{53.85}\% & \SI{82.04}\% \\
    Location and velocity   & 3.9161  & 0.1655 & \SI{62.68}\% & \SI{84.90}\% \\
    Loc., vel., speed, and accel. & 3.6561  & 0.1391 & \SI{64.70}\% & \SI{85.85}\% \\
    \bottomrule
\end{tabular}
\end{table}

\subsection{Effects of Different Context Embeddings} \label{se:ablation_encoders}

In Section~\ref{se:model_ppc}, we utilize three types (PPE, FPE, and FPI) of game context embedding for ball possession prediction. The motivation for including each structure in the model is as follows:
\begin{itemize}
    \item \textbf{PPE:} The necessity of a PPE embedding is straightforward in that the PPC has to be partially permutation-equivariant as explained in Section~\ref{se:pi_pe_encoders}.
    \item \textbf{FPE:} The drawback of only using the PPE embedding is that each player's latent vector $\mathbf{z}_{g,t}^{p}$ does not use the information of opponents. Namely, in Eq.~\ref{eq:ppe_team1}--\ref{eq:ppe_team2}, the model does not consult $(\mathbf{x}_t^{p_{n+1}}, \ldots, \mathbf{x}_t^{p_{2n}})$ when making $\mathbf{z}_{g,t}^{p_1}, \ldots, \mathbf{z}_{g,t}^{p_n}$ and vice versa. Thus, we attach the FPE embedding that covers the information of the entire agents as in Eq.~\ref{eq:fpe}.
    \item \textbf{FPI:} While the above permutation-equivariant embeddings only employ ST-Encoders, a permutation-invariant embedding uses a full Set Transformer including the encoder and the decoder parts. As such, we have intended that the FPI embedding helps to extract additional information about game contexts thanks to the multi-head attention pooling in the decoder part.
\end{itemize}

In addition, we conducted an ablation study to empirically demonstrate that all the variants help improve performance. For each trial, we deployed a subset of these three variants to PPC and measured the performance metrics. The naive ordering model does not use any permutation-equivariant or invariant embedding in PPC. Instead, a single Bi-LSTM directly takes the concatenated feature vectors of roughly ordered (i.e., by uniform number) 22 players as input. Note that the option of only using FPI is impossible for permutation-equivariant encoding since FPI produces a unified output for all the input players.

\begin{table}[htb]
  \caption{Performance comparison between models using different subsets of context embedding variants.}
  \label{tab:ablation_encoders}
  \centering
  \setlength{\tabcolsep}{3.5pt}
  \begin{tabular}{l|ccccc}
    \toprule
   {\bf Embeddings} & {\bf \#. Params} & {\bf PE} &  {\bf RL} & {\bf PPA} & {\bf TPA} \\
    \midrule
    Naive ordering  & 6,860,344 & 7.4223 & 0.3788 & \SI{19.72}{\%} & \SI{72.61}{\%} \\
    PPE         & 7,452,326 & 4.1842 & 0.1938 & \SI{58.56}{\%} & \SI{83.06}{\%} \\
    FPE         & 6,980,630 & 3.8340 & 0.1641 & \SI{57.06}{\%} & \SI{76.82}{\%} \\
    PPE + FPE   & 7,956,662 & 3.7154 & 0.1460 & \SI{64.88}{\%} & \SI{86.16}{\%} \\
    PPE + FPI   & 8,195,510 & 3.8692 & 0.1499 & \SI{62.60}{\%} & \SI{85.14}{\%} \\
    FPE + FPI   & 7,723,814 & 3.8499 & 0.1441 & \SI{56.00}{\%} & \SI{75.68}{\%} \\
    PPE + FPE + FPI & 8,699,846 & 3.6561 & 0.1391 & \SI{64.70}{\%} & \SI{85.85}{\%} \\
    \bottomrule
\end{tabular}
\end{table}

According to the results shown in Table~\ref{tab:ablation_encoders}, the model with all the variants (PPE + FPE + FPI) performs the best. In addition, we can obtain some insights about the role of each embedding as follows:
\begin{itemize}
    \item When not using PPE, PPA, and TPA are significantly degraded. This justifies the use of the PPE embedding that distinguishes the team each player belongs to.
    \item While having strength in predicting ball possession, PPE shows lower performance in ball trajectory prediction than FPE. This is because only using PPE makes the input $\mathbf{\tilde{x}}_t^p$ of the BTR not consult the information of opponents, and this insufficient reference leads to larger position errors for ball trajectory prediction.
    \item As PPE and FPE have their own merits and faults when compared to each other, they create a synergy effect when employed together and show a better performance than when either one is used alone.
    \item FPE and FPI seem to have overlapping roles in that they improve the model performance by a similar amount. Nevertheless, using both of them with PPE (i.e., PPE + FPE + FPI) slightly reduces the position error than using only one (i.e., PPE + FPE or PPE + FPI).
\end{itemize}

\end{document}